\documentclass[useAMS,usenatbib,fleqn]{mn2e}
\usepackage{times}

\usepackage{amsmath}
\usepackage{amssymb}
\usepackage[dvips]{graphicx}

\title[Detecting IGM Scatter Broadening]{Scatter broadening of compact radio sources by the ionized intergalactic medium: Prospects for detection with Space VLBI and the Square Kilometre Array}

\author[J. Y. Koay and J.-P. Macquart]{J. Y. Koay$^{1,2}$\thanks{E-mail:
koayjy@dark-cosmology.dk} and J.-P. Macquart$^{2}$\\
$^{1}$Dark Cosmology Centre, Niels Bohr Institute, University of Copenhagen, Copenhagen \O\, 2100, Denmark\\
$^{2}$International Centre for Radio Astronomy Research, Curtin University, Bentley, WA 6102, Australia\\
\\}

\begin{document}

\date{Accepted 20XX-- Month XX. Received 20XX Month XX; in original form 20XX Month XX}

\pagerange{\pageref{firstpage}--\pageref{lastpage}} \pubyear{2014}

\maketitle

\label{firstpage}

\begin{abstract}

We investigate the feasibility of detecting and probing various components of the ionized intergalactic medium (IGM) and their turbulent properties at radio frequencies through observations of scatter broadening of compact sources. There is a strong case for conducting targeted observations to resolve scatter broadening (where the angular size scales as $\sim \nu^{-2}$) of compact background sources intersected by foreground galaxy haloes and rich clusters of galaxies to probe the turbulence of the ionized gas in these objects, particularly using Space VLBI with baselines of 350,000 km at frequencies below 800 MHz. The sensitivity of the Square Kilometre Array (SKA) allows multifrequency surveys of interstellar scintillation (ISS) of $\sim 100 \,\mu$Jy sources to detect or place very strong constraints on IGM scatter broadening down to $\sim 1\, \mu$as scales at 5 GHz. Scatter broadening in the warm-hot component of the IGM with typical overdensities of $\sim 30$ cannot be detected, even with Space VLBI or ISS, and even if the outer scales of turbulence have an unlikely low value of $\sim 1$ kpc. Nonetheless, intergalatic scatter broadening can be of order $\sim 100\, \mu$as at 1 GHz and $\sim 3\, \mu$as at 5 GHz for outer scales $\sim 1$ kpc, assuming a sufficiently high source redshift that most sight-lines intersect within a virial radius of at least one galaxy halo ($z \gtrsim 0.5$ and $z \gtrsim 1.4$ for $10^{10} {\rm M}_\odot$ and $10^{11} {\rm M}_\odot$ systems, following \citet{mcquinn14}). Both Space VLBI and multiwavelength ISS observations with the SKA can easily test such a scenario, or place strong constraints on the outer scale of the turbulence in such regions.

\end{abstract}


\section{Introduction}\label{introduction}

Baryons consitute only $\sim 4\%$ of the matter-energy budget of the Universe. Of these, only $\sim 10\%$ are found in galaxies, while another 90\% reside in the intergalactic medium (IGM) \citep{fukugitaetal98,fukugitapeebles04}. The baryonic components of the Universe are often classified into four phases based on their temperatures, $T$, and overdensities, $\delta$, relative to the mean baryon density of the Universe \citep{cenostriker99,daveetal01,cenostriker06}, of which the latter three constitute the intergalactic medium (IGM):
\begin{enumerate}
\item \textit{Condensed} -- stars and cool galactic gas residing in galaxies, with $T < 10^5$ K and $\delta > 1000$.
\item \textit{Warm} -- diffuse, photoionized gas, giving rise to Ly$\alpha$ absorption lines in quasar spectra, with $T < 10^5$ K and the distribution of $\delta$ peaked at $\sim 0$. 
\item \textit{Warm-Hot} -- gas shock-heated to temperatures of $10^5 < T < 10^7$ K as they fall into gravitational potential wells to form filamentary structures, 80\% of which have $10 < \delta < 30$.
\item \textit{Hot} -- intracluster gas found in rich clusters of galaxies where large-scale filamentary structures intersect, shock-heated to temperatures of $T > 10^7$ K, and distribution of $\delta$ peaked at $\sim 1000$. 
\end{enumerate}

While both the warm and hot components of the IGM have respectively been detected through Ly$\alpha$ absorption systems (see the review by \citet{rauch98} and the references therein) and in X-ray emission (see review by \citet{sarazin86}), the warm-hot intergalactic medium (WHIM) has been extremely difficult to detect. The WHIM is highly ionized due to the high temperatures, so cannot be detected in absorption except in X-ray lines of heavy elements such as oxygen. It is also too diffuse for its thermal emission to be detected with current X-ray instruments. Cosmological hydrodynamical simulations \citep{cenostriker99,daveetal01,cenostriker06} show that while warm gas consitute $\sim 90\%$ of the mass fraction of baryons at $z \sim 3$, the WHIM becomes the dominant component at $z \sim 0$ with a mass fraction of $\sim 50$\%. Measurements of the mass densities of stars, galaxies and Ly$\alpha$ absorption systems at $z \sim 3$ find that the baryon densities can be accounted for, but summing over all observable contributions at $z \sim 0$ reveal only about 50\% of the baryonic density \citep{fukugitaetal98,fukugitapeebles04}, further evidence that half the baryons reside in this mostly undetected warm-hot phase of the IGM at the present epoch.  

There have been many attempts to search for these `missing baryons' constituting the WHIM at $z \sim 0$ using current UV and X-ray instruments, with various claimed detections of ${\rm O\,\textsc{vi}}$, $\rm O\,\textsc{vii}$ and $\rm O\,\textsc{viii}$ emission and absorption lines associated with the WHIM (see review by \citet{bregman07} and references therein). However, these detections often arise from the tail end of the distribution of WHIM overdensities or temperatures. Additionally, measurements of such highly ionized atoms require further assumptions about the metallicity of the gas in order for total baryon densities to be inferred.  

In the radio regime, various methods have been proposed as to how the WHIM and the ionized IGM as a whole can be detected and studied. \citet{goddardferland03} suggest searching for the hyperfine line of nitrogen $\rm N\,\textsc{vii}$ in absorption, but the atmosphere is opaque at its frequency of occurrence at 53.2 GHz for absorbers at $z \sim 0$. The potential of radio dispersion of impulsive phenomena such as prompt radio emission from gamma-ray bursts to probe the ionized IGM has also been explored \citep{ioka03,inoue04} . Recent detections of fast radio bursts (FRBs) \citep{lorimeretal07,thorntonetal13} have opened up a new avenue for probing the IGM through their radio dispersion \citep{mcquinn14}, found to be in excess of that expected from the Milky Way and thought to originate from the ionized IGM. 

Scattering effects such as scintillation, the temporal smearing of impulsive sources and the angular broadening of compact sources provide another alternative for probing the ionized IGM. These effects, like dispersion, are sensitive not only to the WHIM but all ionized components of the IGM, including the hot intracluster gas and the ionized components of the warm photoionized gas, the latter of which Ly$\alpha$ absorbers trace only the neutral components. In addition to that, scattering also probes the turbulence and density inhomogeneities of these intervening media. \citet{ferraraperna01} and \citet{pallottinietal13} discuss the potential of harnessing intergalactic scintillation to study the IGM. The possibility of detecting the angular broadening of compact sources with the Square Kilometre Array (SKA) is briefly discussed by \citet{lazioetal04}. However, they assume that the scattering mainly occurs in galaxies similar to the Milky Way, rather than in the more diffuse IGM components where the bulk of the baryons reside. Interestingly, two of the 13 FRBs detected to date also exhibit temporal smearing, though whether this scattering originates in the host medium or in the intervening IGM is still hotly debated \citep{kulkarnietal14, luangoldreich14}. In fact, the pulse broadening (and dispersion) of radio bursts from extragalactic supernovae have long been proposed by \citet{meiklecolgate78} as suitable probes of the IGM, while \citet{hallsciama79} have discussed the feasibility of probing angular broadening of compact sources observed through rich clusters of galaxies. The scintillation, angular broadening and pulse broadening of pulsars and other sources due to the interstellar medium (ISM) have been used with great success to map out the distribution of ionized structures in our Galactic ISM \citep{taylorcordes93,gwinnetal93,armstrongetal95,cordeslazio03}, and the aim is to do the same for the IGM with extragalactic sources. 

As the detectability of scintillation and temporal smearing in the IGM have been rigorously discussed elsewhere \citep{pallottinietal13, macquartkoay13}, this paper is solely concerned with the prospects of detecting and probing the various components of the ionized IGM and other turbulent extragalactic structures through the angular broadening of compact sources. Studies of the three effects are complementary since they probe the IGM on different scales and respond differently to the distribution of scattering material along the ray path; angular broadening is preferentially weighted towards material located closer to the observer.  In Section~\ref{model}, we apply the theory of scattering at a thin screen, extended to cosmological scales, to models of various components of the IGM to analyse their contributions to angular broadening. We discuss the strongest limits on angular broadening in the IGM to date (Section~\ref{obconstraints}), based on existing observational data. We then explore and evaluate various strategies for detecting angular broadening with current and next generation radio telescopes in Section~\ref{prospects}, focussing on Space VLBI and the SKA. The main conclusions are summarised in Section~\ref{conclusions}.

\section{Theoretical Estimates and Constraints}\label{model}

In this section, we make use of the thin screen scattering model, extended to cosmological scales, to calculate the angular broadening due to different components of the IGM at different scattering screen redshifts.  

\subsection{Characterizing IGM turbulence and overdensities}\label{turbulence}

A common parameter used to quantify the level of turbulence in the ISM is the spectral coefficient, $C_n^2$, for a truncated power law distribution of the power spectrum of electron density fluctuations in the ISM:
\begin{equation}\label{powerlaw}
P(q) = {C_n^2}q^{-\beta}, \; \frac{2\pi}{l_0} \lesssim q \lesssim \frac{2\pi}{l_1},
\end{equation}
where $q$ is the wavenumber, $l_0$ and $l_1$ are the outer and inner scales of the electron density fluctuations, and $\beta$ is found to have a value of $11/3$ for the ISM \citep{armstrongetal95}, similar to that of Kolmogorov turbulence. In this study, we also adopt the Kolmogorov power-law spectrum of electron density fluctuations for the IGM. While the turbulence properties of the IGM are relatively unknown, there is some observational evidence that intracluster gas exhibit Kolmogorov-like turbulence spectra on tens of kpc scales \citep{schueckeretal04}. The scattering measure (SM), which can be derived from observables, is then the line-of-sight path integral of $C_n^2$ to the source at distance $D_{\rm S}$:
\begin{equation}\label{SMdefinition}
{\rm SM} = \int_0^{D_{\rm S}} ds C_n^2 .
\end{equation}
The SM can also be expressed as \citep{lazioetal08}:
\begin{equation}\label{SM2}
{\rm SM} = C_{\rm SM}\overline{F{n_e^2}}D_{\rm S},
\end{equation}
where the constant $C_{\rm SM} = 1.8 {\rm m}^{-20/3} \, {\rm cm}^6$, $n_e$ is the electron density and $F$ is a fluctuation parameter, given by \citep{taylorcordes93}:
\begin{equation}\label{F}
F = \frac{\zeta \epsilon^{2}}{\eta} {\left( \frac{l_{0}}{1 {\rm pc}}\right)}^{-\frac{2}{3}}. 
\end{equation}
$\zeta$ is the normalized intercloud variance of the mean electron densities of each cloud, $\epsilon$ is the normalized variance of the electron densities within a single scattering cloud, $\eta$ is the filling factor for ionized clouds in the path. 

The SM is a function of the square of the electron densities, which for the IGM will need to be modelled as a function of redshift. The mean free-electron density in the Universe as a function of redshift is given by:
\begin{equation}\label{ne}
n_{e}(z) = x_{e}(z)n_{e,0}(1+z)^3,
\end{equation} 
where $x_{e}(z)$ is the ionization fraction, and $n_{e,0}$ is the mean free electron density at $z \sim 0$. The IGM is known to be significantly ionized out to $z \lesssim 6$ \citep{gunnpeterson65,djorgovskietal01}. Therefore, it can be assumed that $x_{e}(z) \sim 1$ at the redshifts of interest in this study. $n_{e,0} = 2.1 \times 10^{-7} {\rm cm}^{-3}$ assuming hydrogen is fully ionized and helium is singly ionized \citep{yoshidaetal03,inoue04}. At $z \lesssim 3$, helium may be fully ionized, giving $n_{e,0} = 2.2 \times 10^{-7} {\rm cm}^{-3}$ \citep{sokasianetal02,lazioetal08}, which is not significantly different. The latter is used in the present calculations. 

The effects of scattering will be most significant in overdense regions of the IGM, so that Equation~\ref{ne} is modified by an additional term, $\delta_0$, representing the baryon overdensity at $z \sim 0$, giving:
\begin{equation}\label{ne2}
n_{e}(z) = \delta_0 x_{e}(z)n_{e,0}(1+z)^\gamma,
\end{equation} 
where $\delta_0 = p_0/\langle p_0 \rangle$ is the ratio of the baryon density of the IGM component to the mean baryon density of the Universe at the present epoch, and assumes that the electron overdensities are equivalent to the overall baryon overdensities. The manner in which these overdensities scale with redshift will modify the exponent of the $(1+z)$ term, which is quantified here as $\gamma$. For an overdense region that is virialized and gravitationally-bound at the redshifts of interest, as would be expected of the intracluster media (ICM) and galaxy haloes, $\gamma \sim 0$. For components of the IGM that expand with the Hubble flow and have constant comoving densities, $\gamma \sim 3$. In IGM components with $0 < \gamma < 3$, the rate of gravitational infalling into potential wells is lower than the rate of the expansion of the Universe, so that from high to low redshift, the comoving densities are increasing but the proper densities are decreasing. For components with $\gamma < 0$, the rate of gravitational infall is higher than the rate of the expansion of the Universe, so that both the comoving densities and proper densities are increasing from high to low redshift.

\subsection{Scatter broadening at a thin screen at cosmological distances}\label{scattercosmo}

\begin{figure}
     \begin{center}
     \includegraphics[width=70mm]{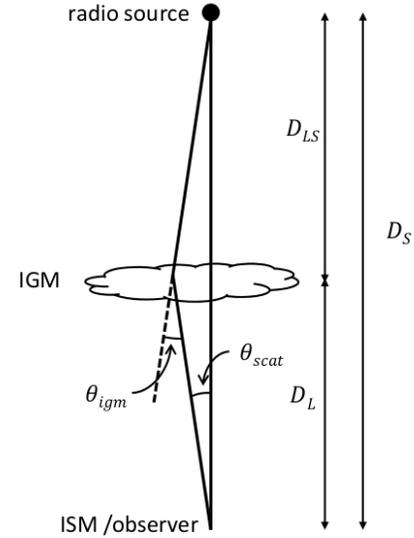}
     \end{center}
     \caption{{The geometry of angular broadening in the intergalactic medium, demonstrating the lever-arm effect. See text in Section~\ref{scattercosmo} for a description of the various symbols used.} \label{leverarm}}
     \end{figure}

To extend thin-screen scattering models to cosmological scales for the IGM, the frequency at the rest frame of the scattering screen needs to be written in terms of the observing frequency, so that $\nu_{\rm screen} = \nu_{\rm obs}(1+z_{\rm L})$ for a screen at redshift $z_{\rm L}$. One must also account for the so called `lever-arm effect' resulting from the geometry of the problem (see Figure~\ref{leverarm}), in converting actual scattering angles in the IGM, $\theta_{\rm igm}$, to the observed scatter broadening angle, $\theta_{\rm scat} = ({D_{\rm LS}}/{\rm D_S})\theta_{\rm igm}$. In the cosmological context, $D_{\rm S}$ is the angular diameter distance to the source and $D_{\rm LS}$ is the angular diameter distance from the source to the scattering screen in the IGM. This follows from the notations used in gravitational lensing literature \citep{narayanbartelmann97,macquart04}. This lever-arm effect is ignored for the scatter broadening of extragalactic sources by the ISM, since $D_{\rm LS} \approx D_{\rm S}$ so that $\theta_{\rm scat} \approx \theta_{\rm ism}$. However, the lever-arm effect has long been recognized to be important for the interstellar scattering of Galactic sources where the distance of the source is finite relative to the screen (see \citet{vandenberg76}, \citet{backer78}, \citet{goodmannarayan89} and \cite{gwinnetal93}). It is also an important consideration for scattering in the IGM, as will be apparent in the following discussion in Section~\ref{modelresults}. 

The size of a scatter broadened image of an extragalactic source due to the IGM with homogenous Kolmogorov turbulence is given in terms of the SM as \citep{macquartkoay13}:
\begin{align}
\theta_{\rm scat} & = \frac{D_{\rm LS}}{D_{\rm S}}\theta_{\rm igm} \label{angIGM}\\ 
              & \sim 19.75 \,\, {\rm SM}^{\frac{3}{5}} \left( \frac{D_{\rm LS}}{D_{\rm S}}\right) {\left( \frac{\nu_{\rm obs}}{1\,{\rm GHz}} \right)}^{-2.2} (1+z_{\rm L})^{-1.2} \,\, {\rm mas}, \notag
\end{align}
for the case where the inner scales of turbulence, $l_1$, are smaller than the diffractive length-scales at the scattering screen, $l_{\rm diff}$. SM is in units of $\rm kpc\, m^{-20/3}$.  Where $l_1 > l_{\rm diff}$, one obtains:
\begin{equation}
\theta_{\rm scat}  \sim 0.31 \,\, {\rm SM}^{\frac{1}{2}} \left( \frac{D_{\rm LS}}{D_{\rm S}}\right) {\left( \frac{\nu_{\rm obs}}{1\,{\rm GHz}} \right)}^{-2} (1+z_{\rm L})^{-1} \left( \dfrac{l_1}{1\,{\rm pc}}\right)\,\, {\rm mas}. \label{angIGM2}
\end{equation}

In the analyses that follow, we focus on the case where $l_1 < l_{\rm diff}$, and note that our conclusions are not significantly affected in the case where $l_1 > l_{\rm diff}$.

\subsection{Model predictions}\label{modelresults}

\begin{figure*}
     \includegraphics[width=\textwidth]{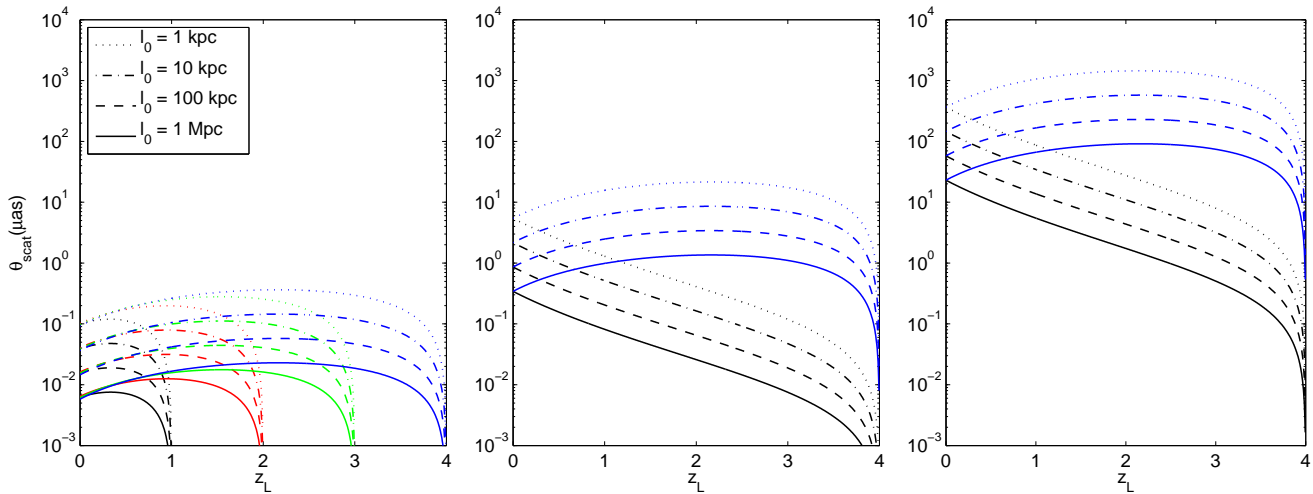}
     \caption{The left panel shows theoretical estimates of angular broadening, $\theta_{\rm scat}$, at 1 GHz due to scattering at a single thin screen at redshift $z_{\rm L}$, with Kolmogorov turbulence cut off at various outer scales, $l_0$. We have assumed the case where the inner scales of turbulence are smaller than the diffractive length-scales at the scattering screen. The electron densities at the scattering screen are assumed to be equivalent to the mean electron densities of the universe at the screen redshift, and therefore have constant comoving densities. The black, red, green and blue curves represent source redshifts of 1, 2, 3 and 4 respectively. The other two panels show similar plots of $\theta_{\rm scat}$ at 1 GHz and source redshift of $z_{\rm S} \sim 4$, but for scattering screens with overdensities of $\delta_0 = 30$ (middle panel) and $\delta_0 = 1000$ (right panel). The blue curves denote screens with constant comoving densities across all redshifts, while the black curves denote screens with constant proper densities.} \label{igmangmod2}
     \end{figure*}

The dependence of $\theta_{\rm scat}$ on the scattering screen redshift is reliant upon two opposing effects. The redshift dependence of the rest-frame frequency at the scattering screen for a fixed observing frequency, together with the geometrical lever-arm effect, causes $\theta_{\rm scat}$ to decrease with increasing $z_{\rm L}$, for a fixed source redshift $z_{\rm S}$. On the other hand, the mean electron density of the Universe scales with $(1+z_{\rm L})^3$, so that the SM scales with $(1+z_{\rm L})^6$ following Equation~\ref{SM2}. The left panel of Figure~\ref{igmangmod2} shows $\theta_{\rm scat}$ for various values of $z_{\rm L}$, $z_{\rm S}$ and $l_0$, for electron densities equivalent to the mean electron densities of the Universe ($\delta_0 \sim 1$ and $\gamma \sim 3$) at all epochs. For simplicity, $\zeta \sim \epsilon \sim \eta \sim 1$ is assumed for all redshifts, implying that the turbulence is fully developed at all redshifts of interest. $l_0$ is also assumed to be independent of $z_{\rm L}$, so that turbulence is continually injected into the IGM at the same scales at all redshifts.  

For IGM components whose densities are coupled to the Hubble flow (and thus have constant comoving densities), angular broadening is strongest for screen redshifts of one half that of the source. At $z_{\rm L} \lesssim 0.5 z_{\rm S}$, $\theta_{\rm scat}$ is dominated by the increasing mean electron densities with redshift, causing $\theta_{\rm scat}$ to increase with $z_{\rm L}$. At $z_{\rm L} \gtrsim 0.5 z_{\rm S}$, the lever-arm effect combined with the increasing rest-frame frequencies of the screen begin to dominate, so that $\theta_{\rm scat}$ decreases with $z_{\rm L}$. The values of $\theta_{\rm scat}$ are of order $\sim 1$ nas to $\sim 0.1 \, \mu$as for outer scales of turbulence between 1 Mpc and 1 kpc, at 1 GHz and for $\delta_0 \sim 1$, with the exception of scattering screens located close to the background source where $\theta_{\rm scat}$ drops to much lower values. This means that scatter broadening is strongly weighted against contributions by the background host galaxy medium, due to the lever arm effect as $D_{\rm LS} \rightarrow 0$. 

$\theta_{\rm scat}$ increases for overdense regions in the IGM, as shown in similar plots in Figure~\ref{igmangmod2}, for a source at $z_{\rm S} \sim 4$ and scattering screens of overdensity $\delta_0 \sim 30$ (middle panel) and $\delta_0 \sim 1000$ (right panel). Screen densities that expand with the Hubble flow are shown as blue curves, whereas gravitationally-bound screens with constant proper densities are shown as black curves. 

Scatter broadening in gravitationally-bound scattering screens is dominated by objects at $z_{\rm L} \sim 0$, since there is no $(1+z_{\rm L})^3$ dependence of electron densities to offset the lever-arm effect and the increasing frequencies at the rest-frame of the scattering screen. Again, scatter broadening in the host medium of the background source is relatively insignificant.  

\subsection{Theoretical constraints}\label{theorconstraints}
The amplitude of the scatter broadening is subject to additional constraints depending on the nature of the turbulence responsible for the underlying electron density fluctuations.  A specific argument advanced by \citet{luangoldreich14} in the context of temporal smearing relates to a model of intergalactic turbulence based on sonic density fluctuations in a diffuse medium.  They argue that the sonic velocity perturbations associated with the density fluctuations deposit energy on a timescale comparable to that on which the waves traverse the outer scale of the turbulence.  This must be large, $l_0 \sim 10^{24}\,$cm $\sim 1\,$Mpc, for the implied heating rate to be comparable to the cooling rate.  

For Kolmogorov turbulence, the amplitude of the density fluctuations on a scale $l$ is $\delta n_e/n_e \sim \left( l/l_0 \right)^{1/3}$ for a medium with a mean density $n_e$.  Following the arguments outlined in \citet{luangoldreich14}, an outer scale of at least $l_0=10^{24}\,$cm ($\sim 1$\,Mpc) therefore implies scatter broadening on an angular scale,
\begin{eqnarray}
\theta_{\rm scat} &\lesssim& \frac{\lambda \, D_{\rm S}^{3/5}}{l_0^{2/5}} \left( \frac{e^2 n_e \lambda}{\pi m_e c^2} \right)^{6/5} \nonumber \\
&=& 3  \, \nu_{\rm GHz}^{-{2.2}} \left( \frac{D_{\rm S}}{1\,{\rm Gpc}} \right)^{3 \over 5} \left( \frac{n_e}{10^{-7}\,{\rm cm}^{-3}} \right)^{6 \over 5}  \,{\rm nas},\nonumber \\
&
\end{eqnarray}
where $e$ is the electron charge, $m_e$ is the electron mass, $\lambda$ is the wavelength of the radiation, $\nu_{\rm GHz}$ its corresponding frequency in GHz, and assuming $l_1 < l_{\rm diff}$. This upper limit is roughly consistent (within a factor of two) with that of the solid lines in the leftmost panel of Figure~\ref{igmangmod2}. This argument may also hold for the WHIM, which would limit scatter broadening in the WHIM to no more than of order $\sim 0.1 \,\mu$as at 1 GHz. A similar limit holds if, conversely, $l_1 > l_{\rm diff}$:
\begin{eqnarray}
\theta_{\rm scat} &\lesssim& \frac{n_e e^2 D_{\rm S}^{1/2} \lambda^2}{l_1^{1/6} \pi m_e c^2 l_0^{1/3}} \nonumber \\
&=& 9 \, \nu_{\rm GHz}^{-2} \left( \frac{D_{\rm S}}{1\,{\rm Gpc}} \right)^{\frac{1}{2}} \left( \frac{l_1}{10^{12}\,{\rm cm}}\right)^{-\frac{1}{6}} \left( \frac{n_e}{10^{-7}\,{\rm cm}^{-3}} \right)  \,{\rm nas},\nonumber \\
&
\end{eqnarray}

However, a counterargument to the applicability of the foregoing limits is that the scattering likely does not occur in such diffuse regions, but instead originates in overdense regions associated with the outer halos of galaxies. Indeed, \citet{mcquinn14} argues that a large fraction of the total intergalactic electron column along any line of sight  emanates from galaxy halos (see Fig.\,1 of \citet{mcquinn14}): a line of sight from a quasar at $z \gtrsim 0.5$ is likely to pass within the virial radius of at least one $10^{10} {\rm M}_\odot$ system, and the line of sight to a $z \gtrsim 1.4$ system is likely to pass within the virial radius of at least one $10^{11}\,{\rm M}_\odot$ system.  The physical constraints implied by the balance of IGM heating and cooling do not apply to such regions.  One might expect a characteristic outer scale of $l_0 \sim 1\,$kpc, which yields  a rough estimate of the magnitude of angular broadening of
\begin{equation}
\theta_{\rm scat} \sim \nonumber
\end{equation}
\begin{eqnarray}\label{thetahaloes}
\begin{cases}
100 \, \nu_{\rm GHz}^{-{2.2}} \left( \frac{D_{\rm S}}{1\,{\rm Gpc}} \right)^{3 \over 5} \left( \frac{n_e}{10^{-4}\,{\rm cm}^{-3}} \right)^{6 \over 5}  \mu{\rm as}, 
	& l_1 < l_{\rm diff},\\
60 \, \nu_{\rm GHz}^{-2} \!\! \left( \frac{D_{\rm S}}{1\,{\rm Gpc}} \right)^{1 \over 2} \!\! \left( \frac{l_1}{10^{12}\,{\rm cm}}\right)^{-\frac{1}{6}} \!\! \left( \frac{n_e}{10^{-4}\,{\rm cm}^{-3}} \right)  \mu{\rm as}, 
	& l_1 > l_{\rm diff},\\
\end{cases} 
\end{eqnarray}
where we normalise to a density of $n_e \sim 10^{-4}\,$cm$^{-3}$ (equivalent to $\delta_0 \sim 1000$) to reflect the fact that the electron density in these regions is expected to far exceed the average density of the diffuse component of the IGM. A similar argument can be applied to intracluster gas, where AGN jets/outflows and galaxy dynamics can inject energy into (or stir) the cluster gas on scales of $\sim 10$ kpc.

\section{Observational Constraints}\label{obconstraints}

\begin{figure*}
     \includegraphics[width=0.8\textwidth]{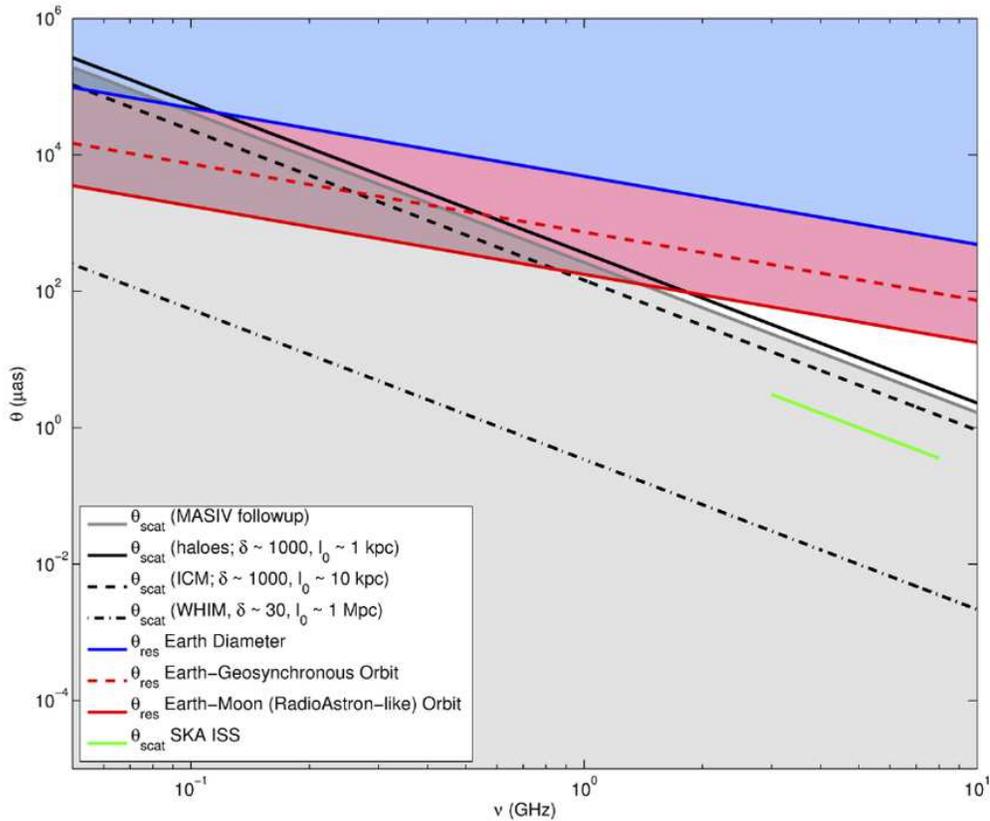}
     \caption{{Observational constraints and model calculations of angular broadening in the IGM, $\theta_{\rm scat}$, shown together with the angular resolutions at various notional array baselines, $\theta_{\rm res}$, extrapolated to various observing frequencies. The gray lines represent the strongest observational upper limits on angular broadening in the IGM to date based on the MASIV follow-up observations \citep{koayetal12}, below which IGM scatter broadening may still be detected towards most sight-lines (gray shaded regions). The blue line represents the highest angular resolution probed by baselines equivalent to the diameter of the Earth, while the blue shaded regions show values of $\theta_{\rm scat}$ that can be probed with ground-based arrays. The red lines show the angular resolutions probed by Space VLBI, with baselines formed between ground stations and stations in geosynchronous orbit (up to $\sim$ 84,000 km), and stations in Moon-like orbits (up to $\sim$ 350,000 km, i.e, RadioAstron or stations on the Moon). The regions shaded in red show the improved range of $\theta_{\rm scat}$ that can be probed by Space VLBI over ground-based VLBI. The model values of $\theta_{\rm scat}$ (black lines) are based on Section~\ref{model} for a screen at $z_{\rm L} \sim 0$ where scatter broadening is expected to be highest for gravitationally-bound components, assuming Kolmogorov turbulence with $l_1 < { l_{\rm diff}}$. We also include in the plot (green line) lower limits of $\theta_{\rm scat}$ that can be probed with multi-wavelength SKA observations of interstellar scintillation (see Section~\ref{ISSresolution}). } \label{igmangconst}}
     \end{figure*}

The strongest constraints for angular broadening in the IGM can be obtained from the most compact extragalactic radio sources known to date --- Gamma-Ray Burst (GRB) afterglows and Active Galactic Nuclei (AGNs) compact enough to exhibit interstellar scintillation (ISS). \citet{frailetal97} estimated the angular size of the source of the radio afterglow of GRB 970508 to be $\lesssim 3\, \mu$as at 8.4 GHz based on observed diffractive ISS of the source. The 5 GHz Microarcsecond Scintillation Induced Variability (MASIV) Survey of 443 compact AGNs found a decrease in the fraction of sources displaying ISS and their scintillation amplitudes above $z \gtrsim 2$ \citep{lovelletal08}, hinting at the possibility of increased angular broadening in the IGM for the high redshift sources. However, dual-frequency follow-up observations of a sub-sample of 128 sources found no significant evidence of scatter broadening in the IGM up to $z \sim 4$ \citep{koayetal12}, constraining the value of $\theta_{\rm scat}$ due to the IGM to $\lesssim 8\, \mu$as at 5 GHz and the SM to $\lesssim 3.3 \times 10^{-5} {\rm kpc \, m}^{-20/3}$ towards the most compact $\sim 10 \,\mu$as objects. This constraint on $\theta_{\rm scat}$ is extrapolated to frequencies ranging from 50 MHz to 10 GHz (gray line) in Figure~\ref{igmangconst}, assuming Kolmogorov turbulence in the IGM where $l_1 < {\rm l_{diff}}$, so that $\theta_{\rm scat} \propto \nu_{\rm obs}^{-2.2}$.  

The observational constraints on scatter broadening in the IGM are broadly consistent with our models and theoretical constraints from Section~\ref{model} for screen redshifts of $z_{\rm L} \sim 0$, shown as black lines in Figure~\ref{igmangconst}. Of particular interest is that the predicted scatter broadening in galaxy haloes appears to be larger than the upper limits from the MASIV follow-up observations. Although this could mean than $l_0$ in these regions are much larger than 1 kpc, or that scatter broadening in the outer regions of galaxy haloes are not significant, the more likely explanation is that the majority of sight-lines towards the most compact sources in the MASIV follow-up observations do not intersect such galaxy haloes at $z \sim 0$. While \cite{mcquinn14} predicts that most sight-lines will intersect within a virial radius of at least one such halo above a redshift of 0.5 or 1.4 (depending on the halo mass), and the MASIV follow-up sources have redshifts between 0 to 4, $\theta_{\rm scat}$ can decrease by a few factors if the intervening halo is located at $z \gtrsim 0.5$ (Figure~\ref{igmangmod2}). This would then decrease the mean $\theta_{\rm scat}$ due to galaxy haloes to that comparable with or slightly below the upper limit imposed by the MASIV observations, for such distributions of background source redshifts. The MASIV follow-up observations are thus close to ruling out or confirming the predictions of \citet{mcquinn14}. Stronger constraints are needed, though the ultimate goal is to retrieve parameters such as $l_0$, $F$, $C_{n}$ and their redshift dependences for various components of the IGM based on actual detections of scattering. 

\section{Prospects for Detection}\label{prospects}

We investigate in this section possible strategies for detecting and probing (or at the very least place even stronger constraints on) scatter broadening in the IGM and other extragalactic structures such as cluster gas and galaxy haloes.

\subsection{Resolving with Space VLBI}\label{directimaging}

One strategy for detecting angular broadening is to resolve compact background sources such as blazars with VLBI at multiple frequencies to determine if their angular sizes scale with $\nu^{-2.2}$, as was attempted by \citet{lazioetal08} without success. In discussing the feasibility of using the SKA to detect intergalactic scatter broadening, \citet{lazioetal04} proposed that angular resolutions better than 4 mas at 1.4 GHz and 80 mas at 0.33 GHz are required. An extrapolation of the constraints from the MASIV follow-up observations gives lower values of $\theta_{\rm scat} \lesssim 126 \, \mu$as at 1.4 GHz and $\theta_{\rm scat} \lesssim 3$ mas at 0.33 GHz. Our model shows that it could be even much lower for the WHIM, with $\theta_{\rm scat} \sim 0.3 \, \mu$as at 1 GHz for $l_0 \sim 1$ Mpc. 

Figure~\ref{igmangconst} demonstrates that even with baselines comparable to the diameter of the Earth (12,700 km), scatter broadening will be barely resolved down to frequencies as low as 50 MHz. This is based only on the strongest observational limits to date, so is true for the majority of sight-lines at least. If the sources intersect regions of overdensities comparable to $\sim 1000$ at $z_{\rm L} \sim 0$, which are more rare, ground-based telescopes may be able to resolve the scatter broadening at frequencies below $\sim 100$ MHz, provided that the outer scales of turbulence in these regions are of order kpc scales as expected for galaxy haloes. 

Space VLBI has the best potential for directly resolving extragalactic scatter broadening. Figure~\ref{igmangconst} shows that for an overdense region of $\delta_0 \gtrsim 1000$ at $z_{\rm L} \sim 0$ with $l_0 \sim 10$ kpc as assumed for intracluster gas, scatter broadening can be resolved below frequencies of $\sim$ 200 MHz and $\sim$ 800 MHz for baselines between ground stations and stations in geosynchronous orbits (up to $\sim$ 84,000 km) and that between ground stations and stations with Moon-like orbits ($\sim$ 350,000 km), respectively. The RadioAstron telescope, which at perigee can form baselines of $\sim$ 350,000 km with a ground based telescope, can thus already place very strong constraints on $l_0$ in cluster gas, provided it can detect sufficiently compact objects behind such structures. It can also test more strongly the suggestion by \citet{mcquinn14} that intergalactic scattering will be dominated by regions close to haloes and will be important for most sight lines above a sufficiently high redshift, or place strong constraints on $l_0$ in such a scenario, for frequencies $< 1$ GHz. However, the contribution of the WHIM with $\delta \lesssim 30$ to angular broadening appears undetectable even with the baselines of RadioAstron. 

Since scattering will be strongest through sight-lines intersecting the most dense and highly turbulent intergalactic structures, we propose targeted observations of compact extragalactic radio sources known to be intersected by galaxy haloes where $\delta \gtrsim 1000$ and $l_0 \lesssim 1$ kpc (perhaps even $\sim$ 1 to 100 pc as observed in the Milky Way \citep{haverkornetal08}) and rich clusters (as suggested by \citet{hallsciama79}). One could envision an experiment using RadioAstron to observe a sample of sufficiently bright compact extragalactic radio sources in the background of a rich cluster of galaxies, or a sample of radio-loud quasars known to be intersected by Mg {\sc ii} and Damped Ly$\alpha$ absorbers to determine if their angular sizes scale with $\nu^{-2.2}$, particularly for intervening objects at $z_{\rm L} \ll z_{\rm S}$. This will allow the turbulence of the ionized medium within these objects, including $l_0$, to be probed or more strongly constrained, particularly since $z_{\rm L}$ can be determined from observations at optical and higher frequencies. These high angular resolution radio observations also provide a good follow-up of candidate detections of WHIM absorption lines at X-ray frequencies to constrain the turbulence in the WHIM, and confirm our predictions. One such WHIM candidate is the detection of an ${\rm O\,\textsc{vii}}$ absorption line at a redshift of $z\sim 0.03$ with estimated overdensities of $\sim 30$ in the X-ray spectrum of the BL Lac object B2356-309 \citep{fangetal10}, with a source redshift $z_s \sim$ 0.1651. Since the search for these X-ray absorption lines tend to focus on BL Lac objects due their featureless spectra, it is likely that these sources will have very compact cores at radio frequencies.

By far the greatest impediment is that angular broadening will be dominated by contributions from the ISM of our Galaxy, the effects of which will be difficult to disentangle from that of the IGM. Even for IGM screens at $z_{\rm L} \sim 0$ where the rest-frame frequency is equal to the observing frequency, the higher density and most likely smaller $l_0$ of the ISM, together with the lever arm effect will favour the ISM. Such observations will therefore need to be carried out off the Galactic plane. Another option is to subtract the Galactic contribution to angular broadening through observations of the angular broadening of pulsars close to the line-of-sight of the target AGN, as has been proposed by \citet{lazioetal04}, or through empirical models of the Galactic electron distribution, e.g. NE2001 \citep{cordeslazio03}. While it is well-known that the NE2001 model has larger uncertainties at higher Galactic latitudes, future pulsar and variability surveys with next generation radio telescopes such as the SKA and its precursors will enable studies of temporal smearing of a larger sample of pulsars, as well as interstellar scintillation of large numbers of both Galactic and extragalactic sources. These will facilitate the construction of more accurate Galactic scattering models, allowing us to better discriminate between Galactic and extragalactic contributions to angular broadening.  

Furthermore, RadioAstron (and Space VLBI in general, due to practical considerations) is limited by its relatively low sensitivity with its 10 m antenna, being able to detect only sources with very high brightness temperature. Studies of IGM scatter broadening with Space VLBI must thus be complemented by alternative methods that can probe angular broadening of fainter background sources, which we discuss next.

\subsection{Resolving with interstellar scintillation}\label{ISSresolution}

Interstellar scintillation (ISS) provides a potent means of detecting IGM scatter broadening in compact AGN. The characteristic frequency dependence of the scintillations can be used to distinguish between the $\nu^{-2.2}$ size dependence characteristic of angular broading and other source size effects intrinsic to the AGN themselves.  ISS is sensitive to the presence (or absence) of structure within a source on angular scales of $\sim 5-100\,\mu$as, and measurement of the amplitude of the intra-day variability at two frequencies is sufficient to ascertain if the source size is dominated by angular broadening.

This method was first implemented by \cite{koayetal12} on a $128$-member sample of compact AGN as a follow-up to the MASIV survey \citep{lovelletal08}.  A summary of the technique, as applied to a sufficiently large sample of AGN that it is ammenable to statistical analysis, is as follows.  

From the point of view of the ISM turbulence responsible for the ISS, the radio sources have an apparent angular size of:
\begin{equation}\label{thetatot}
 \theta_{\rm tot} = \sqrt{\theta_{\rm int}^2 + \theta_{\rm scat}^2} \, ,
\end{equation}
where $\theta_{\rm scat}$ is the scatter broadening contribution by the IGM. We assume that the observations are conducted at sufficiently high Galactic latitudes so that scatter broadening in the ISM before the radio waves arrive at the interstellar scintillation screen is small or negligible. 

The population of background sources is modeled such that they are intrinsically limited to some fixed brightness temperature $T_{\rm b, int} \sim 10^{11}$\,K (e.g.\,due to energy equipartition of the magnetic fields and electrons \citep{readhead94}) so that one has
\begin{align}\label{brightnesstemp}
\theta_{\rm int} &=\sqrt{\dfrac{(1+z_{\rm S}) c^{2} S_{\nu}}{2 \pi \nu^{2} k  T_{\rm b,int} \Gamma}} \propto S_{\nu}^{0.5}\nu^{-1} \nonumber \\
& \sim 3.4\, \mu{\rm as}\, {\left[(1+z_{\rm S}) \left( \dfrac{S_{\nu}}{1\,{\rm mJy}}\right)\right]}^{0.5} \\
& \,\,\,\,\,\,\,\,\,\,\,\,\,\,\cdot {\left( \dfrac{\nu}{5 \,{\rm GHz}}\right)}^{-1} {\left[   \left( \dfrac{T_{\rm{b,int}}}{10^{11}\,{\rm K}}\right)\left( \dfrac{\Gamma}{15}\right)\right]}^{-0.5}   \, , \nonumber 
\end{align}
where $S_{\nu}$ is the observed flux density at observing frequency $\nu$, $\Gamma$ is the Doppler boosting factor due to beaming from bulk relativistic motion within the source, $c$ is the speed of light and $k$ is the Boltzmann constant. 

We introduce the parameter, $R_{\sigma}$, which we define as the ratio of the variance of AGN scintillation amplitudes at two observing frequencies, given as:
\begin{equation}\label{rs}
R_{\sigma} = \dfrac{\sigma_{8.4}^2}{\sigma_{5}^2} \, ,
\end{equation} 
where $\sigma_{8.4}^2$ and $\sigma_{5}^2$ are the variances of the flux density variations exhibited by the source at 8.4\,GHz and 5\,GHz respectively.  We suggest forming a ratio between the variations near these particular frequencies since weak ISS tends to have the largest amplitudes at these frequencies at mid-Galactic latitudes \citep{walker98}; however, the technique is in principle ammenable to any pair of frequencies at which the scintillations can be modelled. 

Since the ISS amplitudes are dependent on the angular sizes of the sources at these two frequencies, $R_{\sigma}$ is dependent on the scaling of source angular size with frequency, and this enables IGM angular broadening effects to be distinguished from intrinsic source size effects (which typically scale as  $\theta_{\rm int} \propto \nu^{-1}$). While ISS is also dependent on other parameters of the ISM such as the scattering screen velocity and the distance between the observer and the scattering screen, these parameters are removed to first order when one calculates $R_{\sigma}$ for each source separately. 

A practical complication arises because the measurements of source variability span a finite duration and are obtained at discrete intervals, so that in practice one uses the amplitude of the flux density structure function on a timescale $\tau$, $D_{\nu}(\tau)$, as a surrogate for the variance; one therefore actually computes
\begin{equation}\label{rd}
R_{\rm D}(\tau) = \dfrac{D_{8.4}(\tau)}{D_{5}(\tau)}\, .
\end{equation} 

\begin{figure*}
     \begin{center}
     \includegraphics[width=\textwidth]{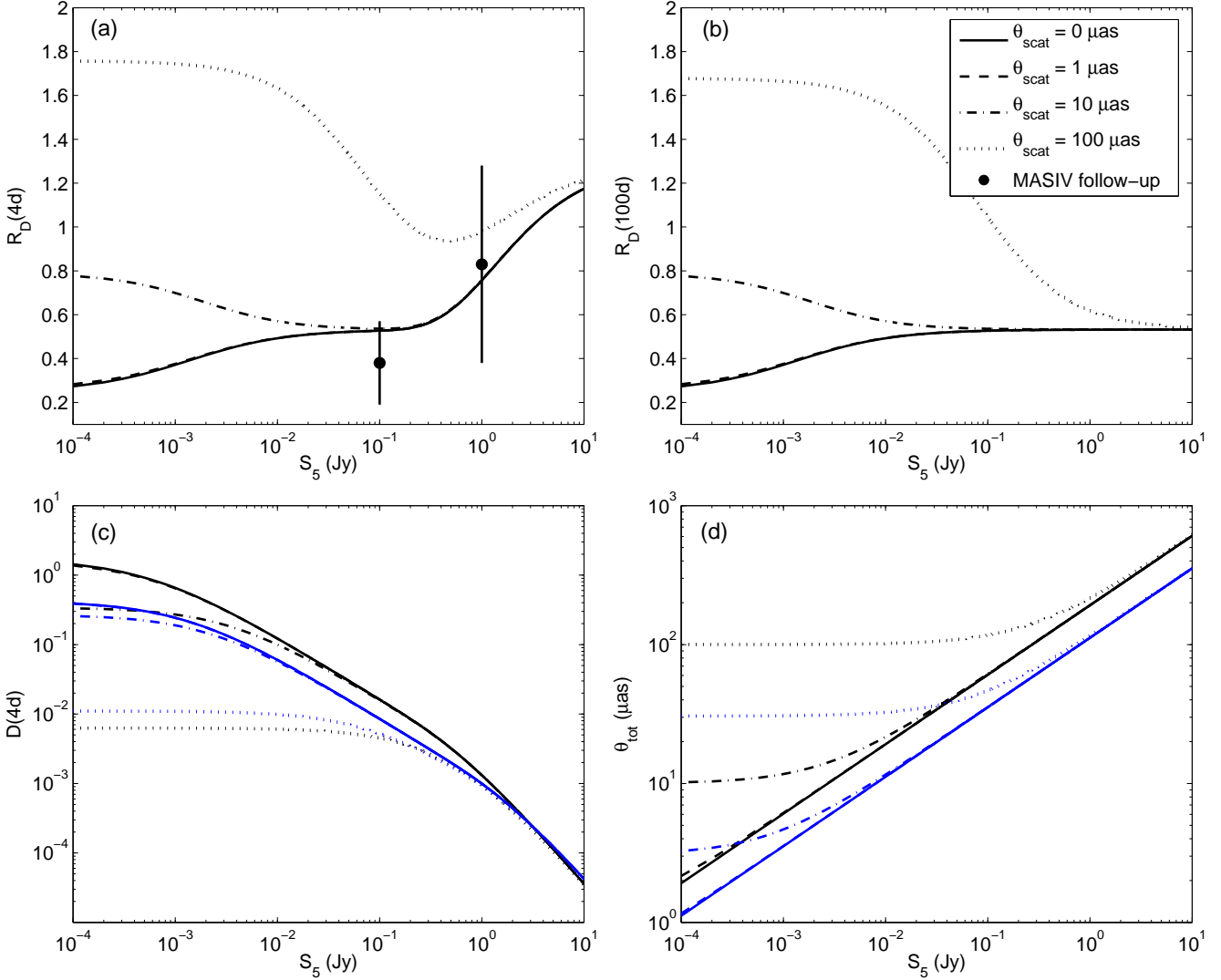}
     \end{center}
     \caption{{(a) Ratio of the 4-day structure function at 8.4 GHz to that at 4.9 GHz, $R_{\rm D}(4{\rm d})$, calculated using the \citet{goodmannarayan06} fitting formula for ISS, versus the mean 5 GHz flux density of the source, for various values of $\theta_{\rm scat}$ at 5 GHz. Also plotted are the fitted $R_{\rm D}$ obtained from the MASIV follow-up observations for $\sim 100$ mJy and $1$ Jy sources \citep{koayetal12}, where the error bars represent the 99\% confidence intervals. (b) Ratio of the 100-day structure function at 8.4 GHz to that at 4.9 GHz, $R_{\rm D}(100{\rm d})$ versus the mean 5 GHz flux density, for various values of $\theta_{\rm scat}$ at 5 GHz. (c) amplitude of the 4-day structure function at 5 GHz (black curves) and 8.4 GHz (blue curves) versus the 5 GHz source mean flux density. (d) The apparent angular source size at 5 GHz (black curves) and 8.4 GHz (blue curves) as seen by the interstellar scintillation screen for various values of the mean 5 GHz flux density. Our models assume that the sources are brightness temperature limited so that the source angular sizes scale with mean flux densities.} \label{ISSresolve}}
     \end{figure*}

One computes the expected value of $R_{\rm D}(\tau)$ based on a given model of the ISM, in conjunction with fitting functions that yield the expected amplitude of ISS, as provided by \citet{goodmannarayan06}.  These functions take into account the scintillation response to the finite angular source size in addition to the expected decrement in variability amplitude between 5 and 8.4\,GHz due to the decrease in scattering strength.  One typically assumes an effective scattering screen distance of $D_{\rm ISM} \sim 500\,$pc, a transition frequency between weak and strong ISS of $\sim 5\,$GHz, and a characteristic scintillation speed of $\sim 50\, {\rm kms^{-1}}$.  However, $R_{\rm D}(\tau)$ is not significantly dependent on some of these parameters of the ISM scattering screen. Figure~\ref{ISSresolve}(a) shows how $R_{\rm D}(\tau = 4\,{\rm days}, 4{\rm d}$) varies with the 5 GHz source flux densities, $S_5$, for different magnitudes of scatter broadening in the IGM at 5 GHz. Corresponding changes in $D_{5}(4{\rm d})$ (black curves) and $D_{8.4}(4{\rm d})$ (blue curves) with $S_{5}$ are also shown in Figure~\ref{ISSresolve}(c). Changes in $\theta_{\rm tot}$ at 5 GHz (black curves) and 8.4 GHz (blue curves) with $S_{5}$ are shown in Figure~\ref{ISSresolve}(d). Here, we have used $\Gamma \sim 15$ and $z \sim 2$, but again these values, although they affect $D_{\nu}(\tau)$ and $\theta_{\rm tot}$ at both frequencies, do not significantly affect $R_{\rm D}(\tau)$ that is computed on a source by source basis, assuming that $\Gamma$ is independent of frequency. 

Typically, when $\theta_{\rm scat} \gg \theta_{\rm int}$, the value of $R_{\rm D}(\tau)$ approaches a constant value $\sim 1.8$, as seen for $\theta_{\rm scat} \sim 100\, \mu{\rm as}$ in Figure~\ref{ISSresolve}(a) when $S_{5} < 10$ mJy ($\theta_{\rm int} < 20\, \mu$as). As $\theta_{\rm int}$ increases with increasing $S_{5}$, $R_{\rm D}(\tau)$ is expected to decrease until it approaches a typical value of $\sim 0.5$ for $\theta_{\rm int} \gg \theta_{\rm scat}$. However, in Figure~\ref{ISSresolve}(a), above $S_{5} \sim 1$ Jy, $R_{\rm D}({\rm 4d})$ increases as the source mean flux density (and $\theta_{\rm int}$) increases. This is because $\theta_{\rm tot}$ is sufficiently large that the characteristic time-scales of ISS are longer than 4 days, so that $D_{\nu}({\rm 4d})$ at both frequencies do not saturate as expected for a stationary stochastic process, with $D_{8.4}({\rm 4d})$ rising faster than $D_{5}({\rm 4d})$ as a function of $\tau$, due to slightly smaller angular sizes and shorter scintillation timescales at 8.4 GHz relative to 5 GHz. This effect can be mitigated by increasing the observational period and using $R_{\rm D}(\tau > 4{\rm d})$ so that the structure functions of the lightcurves at both frequencies saturate, and $R_{\rm D}(\tau)$ is essentially a constant for large values of $S_5$ (as shown in Figure~\ref{ISSresolve}(b) for $R_{\rm D}(100{\rm d}))$.
Alternatively, these larger values of $R_{\rm D}(\tau)$ due to $D_{\nu}({\tau})$ not saturating can be distinguished from large $R_{\rm D}(\tau)$ values typical of sources where scatter broadening dominates by using the observed $D_{\nu}({\tau})$ and $\tau$ at both frequencies as additional model constraints.  

In the analysis of data obtained from the MASIV follow-up observations, the values of $D_{8.4}({\rm 4d})$ and $D_{5}({\rm 4d})$ were calculated at a timescale of $\tau \sim 4\,$days using the model of \citet{goodmannarayan06} and compared with measured values of $R_{\rm D}({\rm 4d})$ based on structure functions measured with the same 4-day time lag. The MASIV observations provided no significant detection of IGM scatter broadening down to the $\sim 10 \,\mu$as level at 5 GHz using a sample of AGN with mean flux densities $\gtrsim S_\nu \sim 100\,$mJy (shown also in Figure~\ref{ISSresolve}(a)). This flux density limit is important, because it means that the MASIV survey and its follow-up observations were constrained to probe relatively large values of $\theta_{\rm int}$, as suggested by Equation~\ref{brightnesstemp}. We note that the error bars for the data points in Figure~\ref{ISSresolve}(a) are 99\% confidence intervals for $R_{\rm D}({\rm 4d})$, obtained from fits to ${D_{8.4}({\rm 4d})}/{D_{5}({\rm 4d})}$ in the paper by \citet{koayetal12}. Only sources at $z \lesssim 2$ were considered, giving a sample of 29 sources at $\sim 100$ mJy and 19 sources at $\sim 1$ Jy, which contributes to the large error bars. 

The observed inverse correlation of ISS amplitudes with mean flux densities in the MASIV Survey \citep{lovelletal08} implies that these scintillating AGN cores indeed may be brightness temperature-limited. Therefore, lower flux density sources tend to have smaller intrinsic source sizes and are more likely to be dominated by scatter broadening. Scatter broadening between $\sim 1$ to $10 \,\mu$as scales is not measurable for sources with flux densities of $S_5 \gtrsim 100\,$mJy, whose angular diameters are dominated by their instrinsic size. The effects of these lower levels of scatter broadening begin to significantly affect $\theta_{\rm tot}$ and $D_{\nu}(4{\rm d})$ at both frequencies for $S_5 \lesssim 1$ mJy ($\theta_{\rm int} \lesssim 6\, \mu$as). Since there is still a factor of $> 2$ increase in the value of $R_{\rm D}$ from $\theta_{\rm scat} \sim 1$ to $\theta_{\rm scat} \sim 10 \,\mu$as, $\theta_{\rm scat}$ can still be probed between these values for sources with flux densities $\lesssim 1$ mJy. For $\theta_{\rm scat} \lesssim 1 \, \mu$as, $R_{\rm D}(4{\rm d})$ no longer deviates significantly from that where the source is not scatter broadened at all, since $\theta_{\rm scat}$ will never dominate at $S_5 \sim 100\, \mu$Jy ($\theta_{\rm int} \sim 2\, \mu$as). 
 
ISS observations of $\sim$ 1 mJy or even $100\,\mu$Jy sources with a highly sensitive instrument such as the SKA has the potential to probe angular sizes at much higher resolution, allowing lower level scatter broadening to be detected. For a full SKA sensitivity of $A_{\rm e}/T_{\rm sys} \sim 12,000\, {\rm m^2 \,K^{-1}}$ as specified by \citet{schillizietal07}, the array would be able to probe $> 5\%$ variations in a 100 $\mu$Jy source at $>5\sigma$ levels with a bandwidth of 350 MHz and an integration time of 1 minute per pointing. Assuming that the SKA will be divided into two subarrays to observe the variability simultaneously at 2 frequencies, $\sim 3$ minutes is required per pointing to achieve these sensitivity levels. 

In fact, the upgraded Jansky Very Large Array (JVLA) can already significantly improve upon what was achieved by the MASIV follow-up observations. With its wider bandwidth, the JVLA can observe $> 5\%$ variations in a 10 mJy source at $>5\sigma$ levels (assuming a 1 GHz bandwidth) with $\sim$ 1 minute integrations per pointing. Again, this assumes the sensitivity of only half the array, with 13 antennas, so that the other half is observing at another frequency simultaneously. This means that a similar experiment to that of the MASIV follow-up observations can be carried out for 10 mJy sources with no increase in observing time (140 sources observed per sidereal day over 11 days). For 5 mJy sources, 3 minutes per pointing is required to achieve the required sensitivity. The flexibility of the WIDAR correlator allows the tuning of two separate 1 GHz bands to two different center frequencies within each receiver band, allowing two sub-arrays to observe ISS at 4 centre frequencies (e.g., 5, 6.5, 8 and 10 GHz), to obtain $R_{\rm D}({\tau})$ from different combinations of frequency pairs for comparisons with the models. Alternatively, each subarray can observe at only a single centre frequency with a contiguous bandwidth of $\sim 2$ GHz to increase the sensitivity and reduce the observing time or observe a larger source sample.   

Observing weaker sources, with $S_5 < 100\,\mu$Jy will not improve the sensitivity towards angular broadening on scales of $\theta_{\rm scat} \sim 1\,\mu$as or lower. The reason is that the apparent angular size of the source, $\theta_{\rm tot}$, as it appears to the scattering screen responsible for ISS drops below the Fresnel scale, the scale at which the scintillations exhibit the effects of a perfectly point-like source in the weak ISS regime, and below which the ISS amplitudes are no longer dependent on the angular size. This is demonstrated by comparing Figures~\ref{ISSresolve}(c) and \ref{ISSresolve}(d) in the cases where $\theta_{\rm scat} = 0$ and 1 $\mu$as. For $S_5 \lesssim 1$ mJy, $D_{\nu}(4{\rm d})$ at both frequencies have saturated even though $\theta_{\rm tot}$ continues to decrease with decreasing $S_5$. $R_{\rm D}(4{\rm d})$ also decreases with decreasing $S_5$ and saturates at $\sim 0.3$ below 1 mJy, and is no longer dependent on the frequency scaling of $\theta_{\rm tot}$, but on the angular size of the Fresnel zone of the ISM scattering screen, which scales as $\propto \nu^{-0.5}$. This is the same reason why $R_{\rm D}(4{\rm d})$ will never approach $\sim 1.8$ for $\theta_{\rm scat} \sim 10 \mu$as, even when scatter broadening completely dominates at $S_5 \lesssim 100\,\mu$Jy. The Fresnel scale is $\sim 5 \, \mu$as at 5 GHz for our chosen screen distance of 500 pc.

The model therefore demonstrates that while such an experiment with a sensitive radio array improves the resolution at which scatter broadening may be probed by weak ISS, the Fresnel scale of the ISM scattering screen places a physical limit below which this is no longer possible. Even though ISS surveys of low flux density sources can place more than an order of magnitude stronger constraints on IGM scattering compared to Space VLBI (Figure~\ref{igmangconst}) at $\sim 5$ GHz frequencies, we are still not able to detect the WHIM at its typical overdensities of $\delta_0 \sim 30$ even if $l_0$ has an unlikely low value of $\sim 1$ kpc. If indeed the contribution to $\theta_{\rm scat}$ from galaxy haloes is important through most lines of sight for sources at sufficiently high $z$, as suggested by \citet{mcquinn14}, instruments such as the SKA will also be able to test this, since Equation~\ref{thetahaloes} gives $\theta_{\rm scat} \sim 3\,\mu$as at 5 GHz for $l_0 \sim 1$ kpc.

There are still advantages to using the SKA for such observations. Its higher sensitivity and survey speeds greatly improves the statistics relative to that of the MASIV Survey and follow-up observations. The SKA will also be able to detect scintillation at a much lower level in the sources, allowing more accurate estimates of $R_{\rm D}({\tau})$ in sources that scintillate less. The main reason for the small source sample used in obtaining $R_{\rm D}({\tau})$ for the MASIV follow-up observations (Figure~\ref{ISSresolve}(a)), is the fact that $R_{\rm D}({\tau})$ can only be estimated to sufficient accuracy when $D_{\nu}({\tau}) > 3\sigma$ above the noise levels at both frequencies. Through proposed synoptic variability surveys \citep{bignalletal14}, the SKA will thus be able to reduce the sizes of the error bars (which for the weak $\mu$Jy and mJy sources will be dominated by errors due to thermal noise and confusion effects in determining $D_{\nu}({\tau})$). Such surveys will also increase the redshifts to which these compact scintillating sources can be detected, as well as the number of sources that can be observed at high redshift, to determine if scatter broadening in the IGM causes a redshift dependence in AGN ISS.  We note in passing that targeted observations of compact radio sources intersected by cluster gas and intervening absorption systems, as proposed in Section~\ref{directimaging}, are equally amenable to the method described here.  In this case, we are probing the $\gtrsim 1$ GHz regimes of scatter broadening in these structures, inaccessible to Space VLBI which can test our models only at frequencies $\lesssim 1$ GHz (see Figure~\ref{igmangconst}).

The disadvantage of the SKA, however, is that many of these compact, flat-spectrum sources will likely contain components that will be resolved by the longest baselines of the array. This results in additional variability as these complex structures rotate relative to the array, increasing the systematic errors. Removing the longest baselines from the SKA when conducting such observations effectively also reduces its sensitivity. Other subsidiary challenges in resolving IGM angular broadening using ISS arise from:
\begin{enumerate}
\item the difficulty of discriminating ISS from intrinsic source variability. This can be done for large samples at a statistical level through correlation with line-of-sight H$\alpha$ intensities. For individual sources however, annual cycles, time-delays between widely separated telescopes, or correlations with variability in the optical and gamma-ray regimes will be required to remove contributions from intrinsic variability.  
\item the limitation of having to carry out these observations at GHz frequencies where ISS is strongest and most rapid at mid-Galactic latitudes (as compared to time-scales of months and years below 1 GHz), but where $\theta_{\rm scat}$ is smaller than that at lower frequencies.  
\item the fact that at the lower end of the flux density scale, below mJy levels, starburst galaxies with more extended radio emission regions begin to dominate the population of radio sources, rather than radio-loud AGNs \citep{seymour08,dezottietal10}.
\end{enumerate}

\section{Summary}\label{conclusions}

In this paper we estimate the angular broadening of a compact continuum radio source due to various components of the IGM for a range of source and scattering screen redshifts. The models are based on an extension of the thin screen scattering model of the ISM to cosmological scales, assuming Kolmogorov turbulence for the IGM. The model predictions were used together with observational constraints obtained from the most compact extragalactic sources known to discuss the feasibility and prospects of detecting scatter broadening in the IGM. The main conclusions can be summarised as follows:

\begin{enumerate}

\item Angular broadening is dominated by the nearest, $z \sim 0$ screens (including the ISM of our galaxy), for scattering screens with fixed proper densities. For scattering screens whose densities are coupled to the Hubble flow, the most efficient scattering occurs at screen redshifts of one half that of the source.

\item Angular broadening in the IGM can barely be resolved with ground-based interferometers, and can only be detected below 800\,MHz at Space VLBI baselines of $\sim 350,000$ km for sight-lines through regions of overdensity $\delta_0 \gtrsim 1000$, for fully developed Kolmogorov turbulence with $l_0 \lesssim 10$ kpc. There is thus a strong case for conducting targeted observations to detect angular broadening of compact background sources intersected by rich clusters of galaxies (as initially proposed by \citet{hallsciama79}) and foreground galaxy haloes (briefly considered by \citet{lazioetal04}) to probe the turbulence of the ICM and ISM of these objects. For the WHIM with $\delta \sim 30$, angular broadening cannot be detected using space VLBI and ISS even if the outer scales of turbulence have an unlikely low value of $\sim 1$ kpc.
\item Multi-frequency surveys of interstellar scintillation of compact extragalactic radio sources with $\sim 100\, \mu$Jy flux densities using the SKA can potentially detect or place even stronger constraints on IGM scatter broadening down to $\sim 1\, \mu$as levels at 5 GHz. 

\item If sight-lines to sources at $z \gtrsim 0.5$ and $z \gtrsim 1.4$ likely intersect regions within a virial radius of a galaxy halo of $10^{10} {\rm M}_\odot$ and $10^{11} {\rm M}_\odot$ respectively \citep{mcquinn14}, angular broadening can be of order $\sim 100\, \mu$as at 1 GHz and $\sim 3\, \mu$as at 5 GHz for $l_0 \sim 1$ kpc through most sight-lines above such redshifts. While this lies close to the strongest observational upper limits to date, both Space VLBI and multi-wavelength ISS observations with the SKA can easily test this scenario, or place strong constraints on $l_0$ of such regions.

\end{enumerate}

\section*{Acknowledgments}

JY Koay is currently supported by a research grant (VKR023371) from Villum Fonden. The Dark Cosmology Centre is funded by the Danish National Research Foundation. Part of this work was carried out while JY Koay was supported by the Curtin Strategic International Research Scholarship (CSIRS) provided by Curtin University. We thank Peter Hall, Dale Frail and Rob Fender for their valuable comments. We also thank the anonymous reviewers for the very helpful and insightful comments that have significantly improved the manuscript.

\bsp

\label{lastpage}


\begin{thebibliography}{99}

\bibitem[Armstrong et al.(1995)]{armstrongetal95} Armstrong J. W., Rickett B. J., Spangler S. R., 1995, ApJ, 443, 209
%
\bibitem[Backer(1978)]{backer78} Backer D. C., 1978, ApJ, 222, L9
%
\bibitem[Bregman(2007)]{bregman07} Bregman J. N., 2007, ARA\&A, 45, 221


\bibitem[Bignall et al.(2014)]{bignalletal14} Bignall H. E. et al., 2014, Advancing Astrophysics with the Square Kilometre Array, in press
%
\bibitem[Cen \& Ostriker(1999)]{cenostriker99} Cen R., Ostriker J. P., 1999, ApJ, 514, 1
%
\bibitem[Cen \& Ostriker(2006)]{cenostriker06} Cen, R., Ostriker, J. P., 2006, ApJ, 650, 560
%
\bibitem[Cordes \& Lazio(2003)]{cordeslazio03} Cordes J. M., Lazio, T. J. W., 2003, arXiv:astro-ph/0207156v3
%

%
\bibitem[Dav\'{e} et al.(2001)]{daveetal01} Dav\'{e} R. et al., 2001, ApJ, 552, 473
%
\bibitem[De Zotti et al.(2010)] {dezottietal10} De Zotti G., Massardi M., Negrello M., Wall J., 2010 A\&AR, 18, 1
%
\bibitem[Djorgovski et al.(2001)]{djorgovskietal01} Djorgovski S. G., Castro S., Stern D., Mahabal, A. A., 2001, ApJ, 560, L5
%
\bibitem[Fang et al.(2010)]{fangetal10} Fang T., Buote D. A., Humphrey P. J., Canizares C. R., Zappacosta L., Maiolino R., Tagliaferri G., Gastaldello F., 2010, ApJ, 714, 1715
%
%
\bibitem[Ferrara \& Perna(2001)] {ferraraperna01} Ferrara A., Perna R., 2001, MNRAS, 325, 1643
%
\bibitem[Frail et al.(1997)]{frailetal97} Frail D. A., Kulkarni S. R., Nicastro L., Feroci M., Taylor G. B., 1997, Nat, 389, 261
%
\bibitem[Fukugita et al.(1998)]{fukugitaetal98} Fukugita M., Hogan C. J., Peebles, P. J. E., 1998, ApJ, 616, 643
%
\bibitem[Fukugita \& Peebles(2004)]{fukugitapeebles04} Fukugita M., Peebles P. J. E., 2004, ApJ, 503, 518
%
\bibitem[Goddard \& Ferland(2003)]{goddardferland03} Goddard W. E., Ferland G. J., 2003, PASP, 115, 647
%
\bibitem[Goodman \& Narayan(1989)]{goodmannarayan89} Goodman J., Narayan R., 1989, MNRAS, 238, 995
%
\bibitem[Goodman \& Narayan(2006)]{goodmannarayan06} Goodman J., Narayan R., 2006, ApJ, 636, 510
%
\bibitem[Gunn \& Peterson(1965)] {gunnpeterson65} Gunn J. E., Peterson B. A., 1965, ApJ, 142, 1633
%
\bibitem[Gwinn et al.(1993)] {gwinnetal93} Gwinn C. R., Bartel N., Cordes J. M., 1993, ApJ, 410, 673
%
\bibitem[Hall \& Sciama(1979)] {hallsciama79} Hall A. N., Sciama D. W., 1979, ApJ, 228, L15

\bibitem[Haverkorn(2008)] {haverkornetal08} Haverkorn M., Brown J. C., Gaensler B. M., McClure-Griffiths N. M., 2008, ApJ, 680, 362
%
\bibitem[Inoue(2004)] {inoue04} Inoue S., 2004, MNRAS, 348, 999
%
\bibitem[Ioka(2003)]{ioka03} Ioka K., 2003, ApJ, 598, L79

%
\bibitem[Koay et al.(2012)]{koayetal12} Koay J. Y. et al., 2012, ApJ, 756, 29
%
\bibitem[Kulkarni et al.(2014)]{kulkarnietal14} Kulkarni S. R., Ofek E. O., Neill J. D., Zheng Z., Juric M. 2014, arXiv/astro-ph:1402.4766

%
\bibitem[Lazio et al.(2004)]{lazioetal04} Lazio T. J. W., Cordes J. M., de Bruyn A. G., Macquart J.-P., 2004, New Astron. Rev., 48, 1439
%
\bibitem[Lazio et al.(2008)]{lazioetal08} Lazio T. J. W., Ohja R., Fey A. L., Kedziora-Chudczer L., Cordes J. M., Jauncey D. L., Lovell J. E. J., 2008, ApJ, 672, 115

\bibitem[Lorimer et al.(2007)]{lorimeretal07} Lorimer D. R., Bailes M., McLaughlin M. A., Narkevic D. J., Crawford F., 2007, Sci, 318, 777

\bibitem[Lovell et al.(2008)]{lovelletal08} Lovell J. E. J. et al., 2008, ApJ, 689, 108
%
\bibitem[Luan \& Goldreich(2014)]{luangoldreich14} Luan J., Goldreich P., 2014, ApJ, 785, L26
%
\bibitem[Macquart(2004)]{macquart04} Macquart J.-P., 2004, A\&A, 422, 761

%
\bibitem[Macquart \& Koay(2013)]{macquartkoay13} Macquart J.-P., Koay J. Y., 2013, ApJ, 776, 125

%
\bibitem[McQuinn(2014)]{mcquinn14} McQuinn M., 2014, ApJL, 780, L33
%
\bibitem[Meikle \& Colgate(1978)]{meiklecolgate78} Meikle W. P. S., Colgate S. A., 1978, ApJ, 220, 1076

%
\bibitem[Narayan \& Bartelmann(1997)]{narayanbartelmann97} Narayan R., Bartelmann M., 1997, arXiv:astro-ph/9606001v2
%

\bibitem[Pallottini et al.(2013)]{pallottinietal13} Pallottini A., Ferrara A., Evoli C., 2013, MNRAS, 434, 3293
%
\bibitem[Rauch(1998)]{rauch98} Rauch M., 1998, ARA\&A, 36, 267

\bibitem[Readhead(1994)]{readhead94} Readhead A. C. S. 1994, ApJ, 426, 51
%
\bibitem[Sarazin(1986)]{sarazin86} Sarazin C. L., 1986, Rev. Modern Phys., 58, 1
%
\bibitem[Schillizi et al.(2007)]{schillizietal07} Schillizi R. T. et al., 2007, SKA Memo, 100
%

\bibitem[Schuecker et al.(2004)]{schueckeretal04} Schuecker P., Finoguenov A., Miniati F., B\"{o}hringer H., Briel U. G., 2004, A\&A, 426, 387

\bibitem[Seymour et al.(2008)]{seymour08} Seymour N. et al., 2008, MNRAS, 386, 1695
%
\bibitem[Sokasian et al.(2002)]{sokasianetal02} Sokasian A., Abel T., Hernquist L., 2002, MNRAS, 332, 601

%
\bibitem[Taylor \& Cordes(1993)]{taylorcordes93} Taylor J. H., Cordes, J. M., 1993, ApJ, 411, 674
%
\bibitem[Thornton et al.(2013)]{thorntonetal13} Thornton D. et al., 2013, Science, 341, 53


\bibitem[Walker(1998)]{walker98} Walker M. A. 1998, MNRAS, 294, 307

%
\bibitem[Vandenberg(1976)]{vandenberg76} Vandenberg N. R., 1976, ApJ, 209, 578
%
\bibitem[Yoshida et al.(2003)]{yoshidaetal03} Yoshida N., Sokasian A., Hernquist L., Springel V., 2003, ApJ, 598, 73
%


\end{thebibliography}
\end{document}